\documentclass[aps,pra,showpacs,twocolumn, superscriptaddress]{revtex4-1}
\usepackage{amsmath}
\usepackage{amsfonts}
\usepackage{braket}
\usepackage{amssymb}
\usepackage{graphicx}
\usepackage{natbib}
\usepackage[colorlinks=true,linkcolor=blue,urlcolor=black,citecolor=blue]{hyperref}
\usepackage{comment}
\usepackage{xcolor}
\definecolor{red}{rgb}{1,0,0}
\definecolor{bostonuniversityred}{rgb}{0.8, 0.0, 0.0}

\newcommand{\da}{\dagger}

\newcommand{\uu}{\uparrow}
\newcommand{\dd}{\downarrow}

\usepackage{lipsum}

\allowdisplaybreaks

\begin{document}
\title{Self-protected adiabatic quantum computation}
\author{P. V. Pyshkin}
\thanks{pavel.pyshkin@gmail.com}
\affiliation{Department of Physical Chemistry, University of the Basque Country UPV/EHU, 48080 Bilbao, Spain}

\author{Da-Wei Luo}
\thanks{davie.hh@gmail.com}
\affiliation{Center for Quantum Science and Engineering and Department of Physics,Stevens Institute of Technology, Hoboken, New Jersey 07030, USA}

\author{Lian-Ao Wu}
\thanks{lianaowu@gmail.com}
\affiliation{Department of Physics, 
University of the Basque Country UPV/EHU, 48080 Bilbao, Spain}
\affiliation{Ikerbasque, Basque Foundation for Science, 48011 Bilbao, Spain}

\date{\today}
\begin{abstract}
Recent experiments show the existence of collective decoherence in quantum systems. We study the possibility of quantum computation in decoherence free subspace which is robust against such kind of decoherence processes. 
This passive protection protocol can be especially advantageous for continuous quantum computation such as quantum annealers.
As an example we propose to use decoherence protected adiabatic quantum computation for the Grover search problem.
Our proposal contains only two-body interactions, making it feasible in near-term quantum devices. 

\end{abstract}

\maketitle

\section{Introduction}
Decoherence is one of the main obstacles to building a scalable quantum computer~\cite{nielsen_chuang_2010}, and is often understood in terms of independent or individual error models.
Recent experiments unexpectedly show that there exists {\em collective decoherence} in our nature~\cite{collective-dephasing-Wilen2021,Collective-2-McEwen2022}.  This newly-discovered phenomenon is triggered by high energy cosmic rays which produce long lifetime phonons in a substrate. These high energy phonons can affect many qubits coherently. 
Thus, it will be important and timely to develop well-tailored protection schemes for quantum computers against such kind of decoherence.

There are several proposals for decoherence suppression and error correction in quantum computation (QC) in the literatures, such as dynamical decoupling~\cite{Viola_BangBang} and quantum error correction~\cite{Shor-9-qbit-code,Steane-error-code}.
In general, one can divide the decoherence suppression and error correction protocols to active ones which involves external pulses, and passive ones which explore symmetry of the system-environment interaction. One of solutions for passive protection against collective decoherence is to process quantum information in decoherence free subspace~(DFS)~\cite{dfs_ref2_1,dfs_ref2_2,dfs_ref2_3,dfs_ref2_4,patent-lidar_wu_blais_2007}.


Different from the gate based QC initially proposed by Deutsch~\cite{Deutsch1989}, where it is possible to carry out error correction procedures easily, it is inconvenient to interrupt continuous QC~\cite{H-oracle-first-Farhi1998} or adiabatic quantum computation (AQC)~\cite{Adiabatic_qc_1} in order to make such corrections. 
Therefore, it is natural to use passive decoherence protection for continuous QC, for instance for the D-Wave system~\cite{Dwave-1-Johnson2011}. In addition,  several self-protection protocols against specific noises have also been proposed for quantum algorithms~\cite{WuByrd} or reported for geometric phases~\cite{Luo2018i,Luo2019d}.

In this paper we show that it is possible to combine DFS either with the continuous QC or the gate based QC.
We propose to illustrate the protocols with the Grover search problem and use Hamiltonians, e.g., directly available for spin chains~\cite{spin-chains-review-1}, as 
there are recent developments in control of spin chains and individual spins~\cite{Spin-Chain-experiment-1-Toskovic2016,Spin-Chain-experiment-2-Baumann2015}, which can be made either in a solid state or generalized directly to trapped ions system~\cite{spin-chains-in-optical-lattice-Simon2011,Spin-Chain-experiment-3-Optical-Neyenhuis2017}. The state-of-the-art technology~\cite{spin-chains-review-1} also allows for preparation of the initial pure quantum state in spin chains, which is essential for a realistic quantum process.

\section{Construction of DFS} 

We consider a system consists of~$n$ qubits (spins). The general form of coupling to the common environment can be described by the Hamiltonian

\begin{equation}\label{tot-ham}
	H = H_s(t)\otimes\mathbb{I} + \mathbb{I}\otimes H_B + Z_t\otimes B,
\end{equation}
where $H_s$ refers to the system, $H_B$ is an environment Hamiltonian, $Z_t = \sum_i^n Z_i$ is a total Z operator, and $B$ is some operator acting in the bath Hilbert space. 
This Hamiltonian can be assumed as an approximation for real dephasing. 

We chose the system Hamiltonian to have some symmetry such that
\begin{equation}\label{ham-commute}
	[H_s(t), Z_t] = 0.
\end{equation}
The next natural assumption is that the initial system-bath state is separable: $\ket{\Psi(0)} = \ket{\psi(0)}\otimes\ket{\chi(0)}$, where $\ket{\psi}$ ($\ket{\chi}$) is system (bath) state. Let us write the evolution governed by~(\ref{tot-ham}):
\begin{multline}\label{glob-evol-1}
	U(T) = \exp{(\displaystyle{-i(\mathbb{I}\otimes H_B + Z_t\otimes B) T})}\times\\
	 \mathcal{T} \exp{(\displaystyle{-i\int_0^T H_s(t')\otimes\mathbb{I}dt'})},
\end{multline}
where~$\mathcal{T}$ is time-ordering operator, and $T$ is evolution time. The factorization of~(\ref{glob-evol-1}) is possible since $[H_s(t)\otimes\mathbb{I}, \mathbb{I}\otimes H_B + Z_t\otimes B ] = 0$, which follows from~(\ref{ham-commute}).
The result of acting~$U(T)$ on the initial state is 
$$U(T)\ket{\Psi(0)} = e^{\displaystyle{-i(\mathbb{I}\otimes H_B + Z_t\otimes B) T}} \ket{\psi(T)}\otimes\ket{\chi(0)},$$ 
where $\ket{\psi(T)} = \mathcal{T} \exp{({-i\int_0^T H_s(t')dt'})} \ket{\psi(0)}$. We assume $\ket{\psi(T)}$ is one of the vectors from computational basis, which means it is also an eigenstate of~$Z_t$. This allows us to write $U(T)\ket{\Psi(0)} = \ket{\psi(T)}\otimes\ket{\chi(T)}$, where $\ket{\chi(T)} = \exp{(-i(H_B + \lambda B) T)} \ket{\chi(0)}$, and $Z_t\ket{\psi(T)} = \lambda\ket{\psi(T)}$. Thus we can see that system and bath remain disentangled.

Let us assume we have an even number of spins. The computational baisis is chosen to be the protected subspace corresponding to $\braket{Z_t}=0$. For example, in the case $n=4$ we have the following basis vectors in this subspace: $\{ \ket{\dd\dd\uu\uu}, \ket{\dd\uu\dd\uu}, \ket{\dd\uu\uu\dd}, \ket{\uu\dd\dd\uu}, \ket{\uu\dd\uu\dd}, \ket{\uu\uu\dd\dd}  \}$. 
It is easy to see that dimension of this subspace is
\begin{equation}\label{subspace_size}
	N\stackrel{\text{def}}{=}\dim{\rm{DFS}} = C_n^{n/2}=\frac{n!}{(\frac{n}{2})!(\frac{n}{2})!}\approx \sqrt{\frac{2}{\pi n}} 2^n.
\end{equation}

\begin{figure} 
	\begin{center}
		\includegraphics{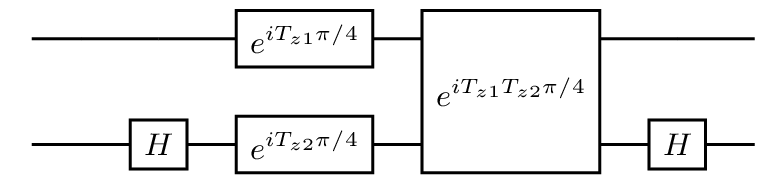}
	\end{center}
	
	\caption{Realization of C-NOT gate in DFS. Each wire corresponds to a logical qubit. Entangling gate realized via Hamiltonian acting on two physical qubits.}
	\label{fig-cnot}
\end{figure}

The simplest system which has a DFS is a pair of spins. In such a case the DFS is $\{ \ket{0}_L, \ket{1}_L \}$, where we denote the states of logical qubit as $\ket{0}_L = \ket{\uparrow\downarrow}$, and $\ket{1}_L = \ket{\downarrow\uparrow}$.
As was shown~\cite{Wu-Lidar-code} we can organize single qubit gates in this subspace with these generators of SU(2) group:
\begin{gather}
	T_x = \frac{X_1X_2 + Y_1Y_2}{2} \label{logical-X}\\
	T_y = \frac{Y_1X_2 - X_1Y_2}{2}\label{logical-Y}\\
	T_z = \frac{Z_1 - Z_2}{2} \label{logical-Z}
\end{gather}
If we consider set of pairs of spins as logical qubits we can rewrite the requirement~(\ref{ham-commute}) in the following way: $[H_l(t), Z_{2l-1} + Z_{2l}]=0$, where $l=1,2,3\dots n/2$ is the number of spin pair, and $H_l(t)$ acts in Hilbert space of this spin pair.
Controlled operation between two logical qubits can be made by $T_{z1}T_{z2} = -Z_2Z_3$, where we assume spins $1$ and $2$ ($3$ and $4$) belong to the first (second) logical qubit~\cite{Wu-Lidar-code}. This allows to us construct Ising gate or more generally~$\exp(iT_{z1}T_{z2}\theta)$ entangling gate acting on a {\em pair of logical qubits} and thus it is possible to realize other two-qubit gates like C-NOT (see Fig.\ref{fig-cnot} and Refs.\cite{Gates-from-H-XX-Makhlin2002,Gates-from-H-XX-Schuch2003}). From the above one can conclude that QC in DFS can be universal and gate-based with known quantum algorithms.

In this paper we propose different models of quantum computation satisfying requirement~(\ref{ham-commute}) which we describe in the next sections.

\section{Continuous Grover search in DFS}

\begin{figure}
	\begin{center}
		\includegraphics{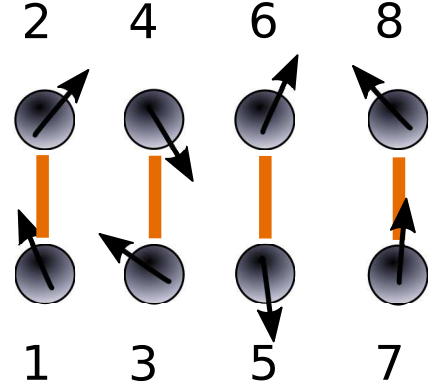}
	\end{center}
	\caption{Independent pairs of spins. Each pair is an antiferromagnetic XXX spin chain.}
	\label{fig_spin-pairs}
\end{figure}

We consider the Grover search problem~\cite{Grover} as an example of using DFS.
Following Farhi et.al.~\cite{H-oracle-first-Farhi1998} we define oracle Hamiltonian as $H_w = -\ket{w}\bra{w}$, where $\ket{w}$ is the unknown state residing in the DFS.
The next step is introducing an equal superposition of all basis vectors in the DFS: 
\begin{equation}\label{s-definition}
\ket{s} = \frac{1}{\sqrt{N}}\sum_{m=1}^N \ket{m}.	
\end{equation}

In order to organize quantum computation we consider the total Hamiltonian which consists of oracle and driving term

\begin{equation}\label{Farhi_Ham}
	H = H_d + H_w = -\ket{s}\bra{s} - \ket{w}\bra{w}.	
\end{equation}
The driving part in~(\ref{Farhi_Ham}) can also be written as
\begin{multline}\label{Farhi_Ham_s}
	H_d = -\ket{s}\bra{s} = \\ - \sum_{k=1}^{n/2} \sum_{m_1<m_2<\dots m_{2k}}^{n} \sigma_{m_1}^\da\dots\sigma_{m_k}^\da\sigma_{m_{k+1}}\dots\sigma_{m_{2k}} + {\rm h.c.}\\ 	
\end{multline}
Here we use~$\sigma_i = \ket{\uparrow}_i\bra{\downarrow}_i$.
The initial state of the system is set to be~$\ket{\psi_0} = \ket{s}$.
The result of unitary evolution governed by~(\ref{Farhi_Ham}) can be written as follows 
\begin{multline}\label{ideal_grover_evolution}
	\ket{\psi(t)} = e^{-iHt}\ket{s} = e^{it}\left\{ \vphantom{\sqrt{1-x^2}}\left(\vphantom{1^1}x\cos(xt) + i \sin(xt)\right)\ket{w} \right.\\
	+ \left.\sqrt{1-x^2}\cos(xt)\ket{r}   \right\},
\end{multline}
where $\ket{r} = (\ket{s} - x\ket{w})/\sqrt{1-x^2}$, and $x = 1/\sqrt{N}$. As can be seen from (\ref{ideal_grover_evolution}) after time~$T=\pi\sqrt{N}$ the state of the system is~$\ket{\psi(T)}\approx\ket{w}$.
It is important to note, $\ket{\psi(t)}$ is evolving in the DFS all the time.

The way we constructed Hamiltonian~(\ref{Farhi_Ham}) on the one hand allows us to use already known results from continuous Grover algorithm, and on the other hand we have a {\em self-protection} of quantum computation from collective dephasing.
Driving Hamiltonian~(\ref{Farhi_Ham_s}) contains many-body interactions which can be hard to realize in practice. Therefore, we propose a more feasible way of QC in DFS in the next section. 

\section{Adiabatic Grover search in DFS}
 
As was discussed above, it is possible to use pair of spins as a logical qubit in DFS, and use existed quantum algorithms with such qubits. 
However, gate based QC requires complicated control such as precise switching off and on interactions, and fields over singles and pairs of logical qubits.

Thus, as a proof of concept we propose to implement adiabatic quantum computation in DFS. In AQC we chose time-dependent system Hamiltonian
\begin{equation}\label{aqc-ham}
	H_s(t) = (1 - s(t)) H_i + s(t)H_f,
\end{equation}
where $H_i$ ($H_f$) are initial (final) Hamiltonian, and $s(0)=1$, $s(T)=1$.
In simple case linear switching $s(t) = t/T$.
Initial state $\ket{\psi(0)}$ is a ground state of the~$H_i$, and in the limit~$T\rightarrow\infty$ the final state~$\ket{\psi(T)}$ is guaranteed to be a ground state of~$H_f$. Obtaining the ground state of~$H_f$ is the goal of AQC.

\begin{figure} 
\begin{center}
	\includegraphics{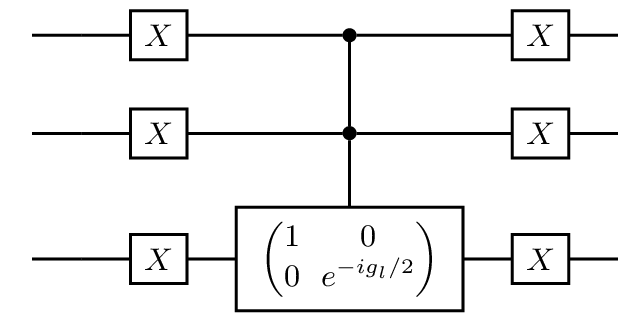}
\end{center}

\caption{Possible realization of the oracle unitary for trotterized AQC in DFS algorithm for $\ket{w}=\ket{000}_L$. Each wire corresponds to logical qubit. Note, this oracle depends on the number of step~$l$ in~(\ref{trotter-U}).}
\label{fig-oracle}
\end{figure}

Let us assume for the moment initial Hamiltonian describes a set of independent XX ferromagnetic spin chains, each of them containing only two spins (Fig.\ref{fig_spin-pairs}). 
Therefore Hamiltonian of each chain is just~$T_x$ from~(\ref{logical-X}).

\begin{figure}
	\begin{center}
		\includegraphics{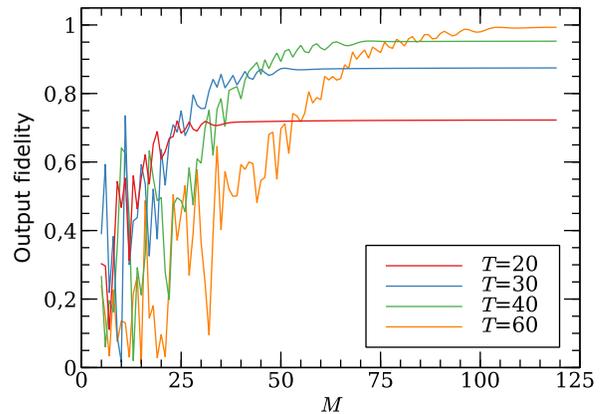}
	\end{center}
	\caption{Fidelity of computation as a function of a number of Trotter steps for different total time of computation. Here we use system of $6$ spins with~$J=1$ and $\dim{\rm DFS}=8$.}
	\label{fig_trotter_steps_1}
\end{figure}

\begin{equation}\label{ham-xx-chain}
	H_i = - \sum_{l=1}^{n/2} \left( \vphantom{\frac{1}{1}}X_{2l-1} X_{2l} + Y_{2l-1} Y_{2l}\right),
\end{equation}
Spectra of a single chain is $\{-2, 0, 0, 2\}$, while there are two states in DFS: $(\ket{\uu\dd} + \ket{\dd\uu})/\sqrt2 = (\ket{0}_L + \ket{1}_L)/\sqrt2$ with eigenenergy $-2$, and $(\ket{\uu\dd} - \ket{\dd\uu})/\sqrt2$ with eigenenergy $2$. Thus we see that ground state of a set of chains is non-degenerate and equal to~$\ket{s}$~(see~(\ref{s-definition})). Moreover, the gap between ground state and the first excited state does not depends on~$n$. 
Assuming $H_f = -\ket{w}\ket{w}$, where $\ket{w}$ is the unknown state from DFS we achieve AQC in DFS which can be described in the same way as was made in~\cite{Adiabatic_qc_1}. However, Hamiltonian~(\ref{ham-xx-chain}) with suppressed~$ZZ$ interaction can be difficult to realize. Nevertheless, we can note that adding $ZZ$ interaction into~(\ref{ham-xx-chain}) does not change the eigenstates.
Moreover, it is well known that antiferromagnetc $XXX$~spin chain is a natural consequence of half-filled Hubbard fermionic model in the limit of strong onsite interaction~\cite{fazekas1999lecture}. 
Thus, instead of~(\ref{ham-xx-chain}) we propose the following initial Hamiltonian    
\begin{equation}\label{ham-xxx-chain}
	H_i =  J\sum_{l=1}^{n/2} \left( \vphantom{\frac{1}{1}}X_{2l-1} X_{2l} + Y_{2l-1} Y_{2l} + Z_{2l-1} Z_{2l}\right),
\end{equation}
where $J>0$. The spectra of a single chain is~$\{-3J, J, J, J\}$, where the non-degenerate ground state is a singlet $|\varphi_g \rangle \equiv (\ket{\uu\dd} - \ket{\dd\uu})/\sqrt2 = (\ket{0}_L - \ket{1}_L)/\sqrt{2}$. 
Note, the Hamiltonian~(\ref{ham-xxx-chain}) in DFS subspace reads $H_L = -J\mathbb{I}_L + 2JX_L$. As we can see the ground of total system is non-degenerate, but it is no longer equal to~$\ket{s}$. We can write the ground state of~(\ref{ham-xxx-chain}) as
\begin{align}\label{GSofXXX}
	\ket{\Psi_0} &= \prod_{n/2}\otimes|\varphi_g \rangle \nonumber\\
	& = N^{-1/2}\left( \sum_{\substack{{\rm number\,of \ket{1}_L} \\ {\rm in\,\ket{n}\,is\, even}}}\ket{n} - \sum_{\substack{{\rm number\,of \ket{1}_L} \\ {\rm in\,\ket{n}\, is\, odd}}}\ket{n}   \right),
\end{align}
where $\ket{n}$ are the states in computational DFS basis like $\ket{0}_L\otimes\ket{0}_L\otimes\ket{1}_L$ e.t.c., and $N = {\rm dim\,DFS} = 2^{n/2}$. Despite the ground state is not an equal superposition like in~\cite{Adiabatic_qc_1} we can use it as initial state in AQC Grover search, due the correspondence given by the following unitary transformation
\begin{equation}\label{GSofXXX}
	U =  \sum_{\substack{{\rm number\,of \ket{1}_L} \\ {\rm in\,\ket{n}\,is\, even}}}\ket{n}\bra{n} - \sum_{\substack{{\rm number\,of \ket{1}_L} \\ {\rm in\,\ket{n}\, is\, odd}}}\ket{n}\bra{n}.
\end{equation}

The oracle Hamiltonian $-\ket{w}\bra{w}$ contains many body interactions like~$Z_1Z_2\dots Z_{n/2}$ which usually doesn't appear in nature. Thus, for the proof of concept purpose we propose to realize AQC in DFS via Trotterization~\cite{Trotterization-Lloyd1996,Wu2002-trotterization,Trotterization-Smith2019,Trotterization-Tacchino2020}, i.e. dividing evolution time $T$ to small parts $\Delta t = T/M$ with $M\gg1$ and approximate smooth adiabatic evolution by a sequence   
\begin{equation}\label{trotter-U}
	U(T)\approx \prod_{l=1}^M \left( e^{-iH_fg_l/2K}e^{-iH_if_l/K}e^{-iH_fg_l/2K}\right)^K,
\end{equation}
where for the linear switching~(\ref{aqc-ham}) $f_l = (1 - \Delta t\cdot l / T)\Delta t$, $g_l = \Delta t^2\cdot l / T $. 
Such a decomposition is valid when $\Delta t|| H_i - H_f||\ll1$. Also, parameter~$K$ should be big enough $K\gg M\Delta t^3$ (see~\cite{Trotterization-Tacchino2020}). However, in our numerical examples we use~$K\gtrsim 1$, which provides relatively good output fidelity for small systems, and reduces number of gates, which is crucial for the proof of concept implementation on near-future quantum computers.

\begin{figure}
	\begin{center}
		\includegraphics{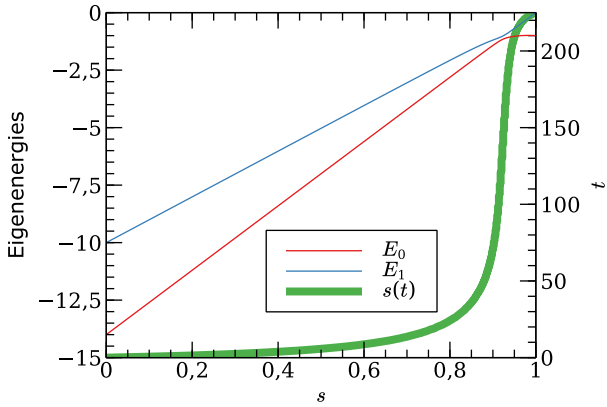}
	\end{center}
	\caption{Two lowest eigenenergies of Hamiltonian~(\ref{aqc-ham}) and optimized adiabatic path~$s(t)$.}
	\label{fig_optimized_path}
\end{figure}

Each Trotter step consists of two different types of unitary operations governed by~$H_i$ and~$H_f$ Hamiltonian. It is already known how to implement oracle in circuit model. Thus, the operation $e^{-iH_fg_l/2}$ (which is {\em phase oracle}) can be made with using set of gates in DFS~(\ref{logical-X},\ref{logical-Y},\ref{logical-Z}) and control-Z\cite{Wu-Lidar-code} (see Refs.\cite{nielsen_chuang_2010, 3-qubit-oracles} and Fig.\ref{fig-oracle} for scheme of the oracle). Operation $e^{-iH_if_l/2}$ is just an evolution governed by Hamiltonian~(\ref{ham-xxx-chain}) and does not require implementation additional gates like Grover diffusion operator in circuit model. Thus, despite of using Trotterization and unitary gates instead of direct realization of oracle Hamiltonian, proposed method has an advantage over circuit search model, because there is no need to organize set of gates for diffusion operator. This can help to avoid control errors.
As first numerical example we use a system of $6$ spins, which corresponds to $3$ logical qubits with $\dim{\rm DFS} = 8$. In Fig.\ref{fig_trotter_steps_1} we show the resulted fidelity of computation as a function of the number of steps~$M$ for different times~$T=20,30,40,60$. Saturation of curves correspond to continuous adiabatic evolution governed by~(\ref{aqc-ham}).

\begin{figure}
	\begin{center}
		\includegraphics{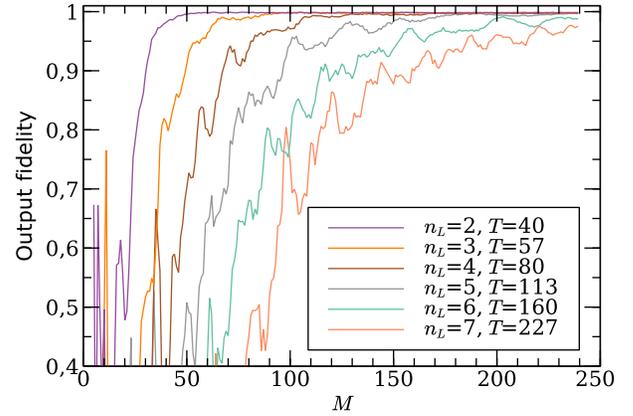}
	\end{center}
	\caption{Fidelity $|\braket{w|U(T)|\psi(0)}|^2$ of computation as a function of a number of steps~$M$ for different total time of computation and different size of the system ($n_L=n/2$ is the number of logical qubits). Here we use optimized adiabatic switching with Trotter parameter $K=1$.}
	\label{fig_trotter_steps_2}
\end{figure}

\begin{figure}
	\begin{center}
		\includegraphics{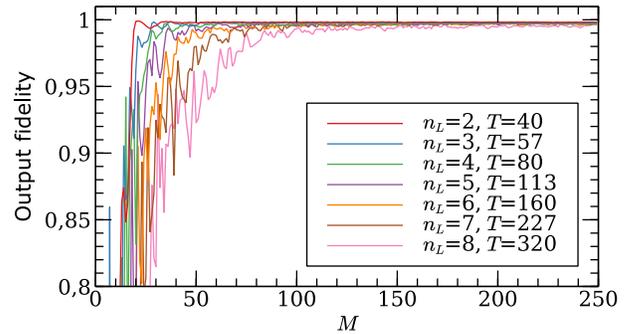}
	\end{center}
	\caption{Fidelity $|\braket{w|U(T)|\psi(0)}|^2$ of computation as a function of a number of steps~$M$ for different total time of computation and different size of the system ($n_L=n/2$ is the number of logical qubits). Here we use optimized adiabatic switching with Trotter parameter $K=n_L$.}
	\label{fig_trotter_steps_3}
\end{figure}

Linear switching in~(\ref{aqc-ham}) is not optimal and does not provide a quantum speedup~\cite{Adiabatic_qc_1}. Following~\cite{Adiabatic_qc_2} we can improve efficiency by using the following changes in~(\ref{trotter-U}): $f_l = (1-s)\Delta t$, $g_l = s\Delta t$, where for each step~$l$ we can find corresponded value of~$s$ by solving the following equation
\begin{equation}\label{optimized_1}
	l\Delta t = T \dfrac{ \int_0^s \dfrac{ds'}{ (E_1(s') - E_0(s'))^{2}} } {  \int_0^1 \dfrac{ds'}{ (E_1(s') - E_0(s'))^{2}} },
\end{equation} 
where $E_{0,1}(s')$ are ground and next after eigenenergies of Hamiltonian $H(s')$~(\ref{aqc-ham}). Expression~(\ref{optimized_1}) written in such a way to satisfy~$M\Delta t = T$. In Fig.~\ref{fig_optimized_path} we show example of dependence~$E_{0,1}(s)$ and $s(t)$ for the case $n=14$ and $T=225$ ($\dim{\rm DFS}=2^7$). In Fig.~\ref{fig_trotter_steps_2} we show numerical simulation with using optimized switching from $H_i$ to $H_f$. As can be seen from Fig.~\ref{fig_trotter_steps_2}, it is enough $K=1$ in (\ref{trotter-U}) for system with $<5$ logical qubits. The increasing Trotter parameter~$K$ improves output fidelity. In Fig.~\ref{fig_trotter_steps_2} we show output fidelity with $K=n_L$, where $n_L=n/2$ is the number of logical qubits. In both figures~\ref{fig_trotter_steps_2} and~\ref{fig_trotter_steps_3} we chose evolution time to increase by a factor of $\sqrt{2}$, i.e. $T\rightarrow\sqrt{2}T$ for each increment of the number of logical qubits $n_L\rightarrow n_L+1$.

\begin{figure}
    \centering
    \includegraphics[scale=.78]{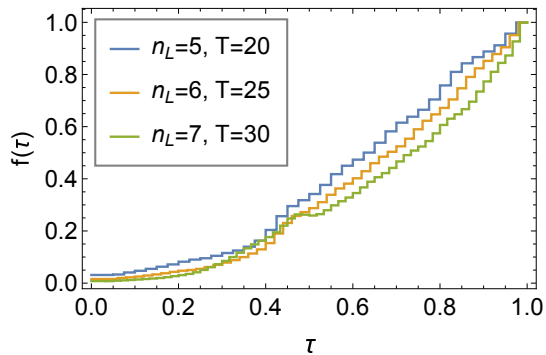}
    \caption{Using the Krotov method to optimize the switch function, we show the output fidelity $f$ as a function of the scaled time $\tau = t/T$, for various number logic qubits. The final output fidelity is $\{0.999683, 0.999368, 0.999069\}$ up from $\{0.854497, 0.789701, 0.691594\}$ using the vanilla switching function obtained from Eq.~\eqref{optimized_1}, for $n_L = 5,6,7$ respectively. Here we have used $M=2T$ and Trotter parameter $K=1$.}\label{fig_krotov_f}
\end{figure}

\begin{figure}
    \centering
    \includegraphics[scale=.78]{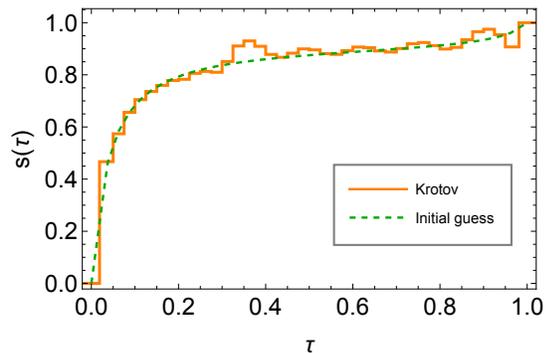}
    \caption{The switching function $s(\tau)$ as a function of the scaled time $\tau=t/T$ for $n_L=5$ logic quibits, parameters are the same as Fig.~\ref{fig_krotov_f}. The orange line is the optimized result using the Krotov method, and the dashed green line is obtained via Eq.~\eqref{optimized_1} which serves as an initial guess for the Krotov method.}\label{fig_krotov_s5}
\end{figure}

In addition to increasing the Trotter steps $M$ or parameter $K$, both of which increase the number of gates required, it is also possible to increase the output fidelity by further optimizing the switching function $s(t)$. One such way to optimize the switching function is by means of the Krotov method~\cite{Sklarz2002r,krotov-book,Zhang2016f}. The switching function obtained from Eq.~\eqref{optimized_1} can be used as an initial guess for the Krotov method. Since the monotonic convergence of the Krotov method is only guaranteed in the continuous control limit, the parameters for the Krotov method have also been appropriately chosen to account for the coarse time step. In Fig.~\ref{fig_krotov_f} we show the fidelity as a function of the scaled time $\tau=t/T$ for $n_L=5,6,7$ logic qubits. We can see that the final output fidelity can all reach $0.999$, which is a quite noticeable increase compared with the vanilla switching function Eq.~\eqref{optimized_1} under the same parameters. It's worth pointing out that this is achieved by optimizing the switching function under a relatively short runtime and small number of Trotter steps, which reduces the number of gates required to carry out the search algorithm. In Fig.~\ref{fig_krotov_s5}, we show the optimized switching function and the vanilla switching function obtained from Eq.~\eqref{optimized_1} as a function of the scaled time $\tau=t/T$ for $5$ logical qubits. It can be seen that the two agree well initially, when the energy gap between the two lowest eigenenergies are large, and small corrections are made when the energy gap becomes smaller.

\section{Conclusions} 

We propose to use DFS as a computational space for continuous QC. 
At first we show that the Farhi~\cite{H-oracle-first-Farhi1998, Adiabatic_qc_1} proposals of continuous and adiabatic QC can be implemented in DFS.
Next, we show that it is possible to achieve QC in DFS via both continuous and gate-based where each logical protected qubit consists of two physical qubits with XXX interaction between each other.  
This passive protection does not require application of complicated external pulses. Also, to realize this protection only two-body interactions is necessary.

\section*{Acknowledgments}
We thank E.~Ya.~Sherman for helpful discussions. We acknowledge support of the Spanish Ministry of Science and the European Regional Development
Fund through PGC2018-101355-B-I00 (MCIU/AEI/FEDER, UE), and the Basque Country Government through
Grant No. IT986-16.


\end{document}